\documentclass[a4paper,fleqn,usenatbib]{mnras}

\usepackage[T1]{fontenc}
\usepackage{ae,aecompl}

\usepackage{graphicx}	
\usepackage{amsmath}	
\usepackage{amssymb}	

\title[Faraday screen surrounding Coma A]{The magnetic field strength of the Faraday screen surrounding the radio galaxy Coma A}

\author[S. Knuettel et al.]{
S. Knuettel,$^{1}$\thanks{E-mail: s.knuettel@mars.ucc.ie}
S. P. O'Sullivan,$^{2}$
S. Curiel$^{3}$
and B.H.C. Emonts$^{4}$
\\
$^{1}$Department of Physics, University College Cork, Cork, Ireland\\
$^{2}$Hamburger Sternwarte, Universit\"at Hamburg, Gojenbergsweg 112, 21029 Hamburg, Germany;\\
$^{3}$Instituto de Astronom\'{i}a, Universidad Nacional Aut\'{o}noma de M\'{e}xico (UNAM), A.P. 70-264, 04510 M\'{e}xico, D.F., Mexico\\
$^{4}$National Radio Astronomy Observatory, 520 Edgemont Road, Charlottesville, VA 22903, USA
}

\date{Accepted 2018 November 03. Received 2018 October 30; in original form 2018 August 01}

\pubyear{2018}

\begin{document}
\label{firstpage}
\pagerange{\pageref{firstpage}--\pageref{lastpage}}
\maketitle

\begin{abstract}
Studying the interaction between AGN jets and lobes and their surrounding environment is important in order to understand how they transfer energy to their environment as well as determining the intrinsic physical properties of the sources themselves. This paper presents broadband VLA polarization and Faraday rotation observations of the radio galaxy Coma A (3C\,277.3) from 1 to 4~GHz, including archival VLA observations at 4.9 and 15~GHz. Through broadband polarization model-fitting, we find that an external Faraday screen with a turbulent magnetic field provides an appropriate description to most of the data. By combining the polarization and Faraday rotation results with previous H\,$\alpha$ observations, we identified the H\,$\alpha$-emitting gas as the Faraday screen responsible for the observed Faraday depolarization. We were able to derive the magnetic field strength in the H\,$\alpha$-emitting gas, finding typical field strengths of $\sim1$~$\umu$G, which is consistent with studies of the intra-group medium local to other radio galaxies. However, we find a highly depolarized region of the southern lobe coincident with a H\,$\alpha$ filament that has a field strength comparable to the equipartition field strength in the radio lobe (i.e.~$\gtrsim$36~$\umu$G). This implies that the H\,$\alpha$ filament is internal to the radio emitting plasma. Such clear examples of internal Faraday depolarization are rare, thus providing another key insight into the evolution of radio galaxies and their ability to provide significant feedback on the local gas that would otherwise cool and form stars.  
\end{abstract}

\begin{keywords}
galaxies: magnetic fields -- galaxies: active -- galaxies: individual: Coma A -- radio continuum: galaxies -- techniques: polarimetric
\end{keywords}




\section{Introduction}
    The radio-emitting structures in the vicinity of supermassive black holes in the centres of active galaxies display a large variety of morphological properties. It is a long standing debate to what extent the differences between different types of radio sources are due to intrinsic properties of the central engine (`nature') or the properties of the interstellar and intergalactic medium (ISM and IGM) surrounding the central engine and host galaxy (`nurture'). Investigating this `nature vs. nurture' debate for nearby radio galaxies is vital for understanding the properties and evolution of radio galaxies throughout the Universe \citep[e.g.][and references therein]{Fabian2012,Heckman2014}.
    
     \par Here we study one such case, where there are clear interactions between the radio lobes and the surrounding gas, in the nearby radio galaxy Coma A \citep{VanBreugel1985,Bridle1981,Tadhunter2000,Morganti2002,Worrall2016}. 
     In particular, we present a study of the combination of wide-band polarimetric radio observations and deep H\,$\alpha$ images. This combination provides a powerful tool to investigate the interactions between the lobes and their surrounding environment: the wide-band radio observations allow us to investigate the polarization and magnetic field structure of the radio emitting lobes and how this changes with wavelength, and the H\,$\alpha$ measurements allow estimation of the electron density in the shocked gas surrounding the lobes, which in turn enables the calculation of the magnetic field strength in the gas containing the free electrons using the effect of Faraday rotation.

    \subsection{Coma A}
    \label{subsec:known_info}
    Coma A is a radio galaxy at redshift $z=0.0857$ \citep{Bridle1981} corresponding to a comoving radial distance of 371.6~Mpc and angular scale of  1.659~kpc\,/\,\arcsec~ \citep{Wright2006} using the values of $H_0= 67.74\,\rm{km \, s}^{-1}\rm{Mpc}^{-1}$, $\Omega_M = 0.3089$, and $\Omega_\Lambda = 0.6911$ from the Plank 2015 results \citep{Planckcoll2016}.  The source has two distinct radio lobes and an angular size of 45\arcsec~ in the north-south direction. Coma A is classified as a Fanaroff-Riley II galaxy in \cite{Fanaroff1974}. The northern lobe shows a bright hotspot, and the southern lobe has a bright plume expanding into a lobe structure. This plume is at a deflected angle to the expected jet direction due to a collision with a gas cloud, decelerating the jet and giving it the features of an FRI sturcture \citep{VanBreugel1985}. Due to this feature and the fact that the power of this radio galaxy is lower that most FRII galaxies, Coma A is on the border between FRI and II classification \citep[][Table~1]{Fanaroff1974}. It was classified as an FRI for this reason in \cite{Chiaberge1999}. It is an example of a hybrid morphology radio source \citep[HyMoRS][]{Gopal-Krishna2000b,Gawronski2006,DeGasperin2017}.  
    
    \par Large scale arcs and filaments of H\,$\alpha$ emission have been observed surrounding the radio lobes \citep{Tadhunter2000}, which closely match the shape of the radio lobes of Coma A. The origin of this ionised gas is discussed in depth in \cite{Tadhunter2000}. The most likely cause for the ionisation is the shock from the expanding radio lobes moving into a region of neutral gas surrounding the host galaxy. The origin of the gas has been investigated with HI absorption observations of Coma A. These show the presence of HI local to both lobes and suggest that a large scale gas disk surrounds the galaxy, into which the radio lobes are expanding \citep{Morganti2002}. The origin of the gas itself is most likely from a merger in the Coma A host galaxy. The existence of such gas around the lobes is also observed in the X-ray. In particular, \citet[][Figure~7]{Worrall2016} shows how regions of diffuse X-rays in the local environment display an anti-correlation to the H\,$\alpha$ filaments. They argue that the original group atmosphere has been shock ionised and now surrounds the lobes, causing the extended X-rays and the H\,$\alpha$ filaments.
   \par In this paper, we present new wide-band dual-polarization observations of Coma A using the VLA. We use these and archival VLA observations to investigate how the H\,$\alpha$-emitting gas and radio lobes interact, through an analysis of the Faraday rotation of the polarized radio emission.

    \subsection{Polarization and Faraday rotation}
    \label{subsec:RM_QU_theory}
    
    Radio galaxy lobes emit synchrotron radiation from relativistic electrons accelerated by magnetic fields. The observed flux density typically obeys a power law with frequency, $ S \propto \nu^{+\alpha}$, where $S$ is the observed radio flux density, $\nu$ is the observing frequency and $\alpha$ is the spectral index. Synchrotron radiation is intrinsically highly linearly polarized, with a theoretical maximum limit of $\sim 70\%$. The linear polarization is measured using the Stokes parameters, $Q$ and $U$ where the complex linear polarization ($P$) is defined as    
    \begin{equation}
    P = Q + \mathrm{i}U = p_0 I \mathrm{e}^{2\mathrm{i}\chi}
    \label{eq:complex_lin_pol}
    \end{equation}    
    where $p_0$ is the intrinsic degree of polarization (i.e at zero wavelength), $\chi$ is the observed polarization angle and $I$ is the total intensity. The fractional polarization, $p$, and polarization angle, $\chi$, are described as follows.
    \begin{equation}
    p=\frac{\left|P\right|}{I}=\dfrac{\sqrt{Q^2 +U^2}}{I}
    \label{eq:frac_pol_eq}
    \end{equation}
    \begin{equation}
    \chi= \frac{1}{2}\arctan\left(\frac{U}{Q}\right)
    \end{equation}
    The observed polarization angle, $\chi$, changes with wavelength due to Faraday rotation. This effect acts on a polarized electromagnetic wave as it passes through a region with free electrons and magnetic field along the line of sight. The polarization angle changes linearly as a function of wavelength squared in the simplest case,    
    \begin{equation}
    \chi = \chi_0 + \mathrm{RM}\lambda^2
    \label{eq:basic_chi}
    \end{equation}   
    where $\chi_0$ is the intrinsic polarization angle and RM is the Faraday rotation measure. This RM value is calculated as follows \citep{Burn1966}:    
    \begin{equation}
    \label{eq:Burn_RM_eqn}
    \mathrm{RM} = 0.81 \int n_e\left(l\right) B_\parallel \mathrm{d}l 
    \end{equation}   
    Where $B_\parallel$ is the line of sight component of the magnetic field in $\umu$G, $n_e$ is the electron density in cm$^{-3}$, and $\mathrm{dl}$ is an element along the path along the line of sight in parsecs.
    \par It is commonly observed that the fractional polarization, $p$, tends to decrease towards longer wavelengths, an effect known as depolarization. One of the most common models associated with this phenomenon is external Faraday dispersion \citep{sokoloff1998}, where an external screen with free electrons and a turbulent magnetic field depolarizes the signal within the observing beam, due to the different resulting Faraday rotations occurring along individual lines of sight within the beam. It is possible to fit models to observed data with good $\lambda^2$ coverage in order to estimate the parameters of the source polarization and turbulent screen \citep[e.g.][]{OSullivan2012}. The model used in this paper has a single RM component and an external depolarizing screen, motivated in part by the picture of a shell of ionized gas surrounding the lobes of Coma A:
    \begin{equation}
    \label{eq:model_eqn}
    p\left(\lambda\right) = p_0 \mathrm{e}^{2\mathrm{i}\left(\chi_0 + \mathrm{RM}\lambda^2\right)}\mathrm{e}^{-2\sigma_\mathrm{RM}^2\lambda^4}
    \end{equation}
    \noindent Here, $p\left(\lambda\right)$ is the fractional polarization as a function of wavelength. The first exponential term is the complex linear polarization (Equation \ref{eq:complex_lin_pol}) with the observed polarization angle defined as in Equation \ref{eq:basic_chi}.
    The second exponential term represents the depolarization by an external Faraday screen where $\sigma_{\mathrm{RM}}$ is the standard deviation of the RM within the telescope beam, due to the turbulent magnetic field in the ionized gas. 
    Using Eqns.~\ref{eq:complex_lin_pol} and \ref{eq:model_eqn}, the fractional Stokes $Q$ and $U$ parameters, $q(\lambda)$ and $u(\lambda)$, are described as follows.
    \begin{subequations}
    \begin{align}
    q\left(\lambda\right) = p_0 \cos\left(2\chi_0 + 2\mathrm{RM}\lambda^2\right) \mathrm{e}^{-2\sigma^2_{\mathrm{RM}}\lambda^4} \label{eq:modelfitQ} \\
    u\left(\lambda\right) = p_0 \sin\left(2\chi_0 + 2\mathrm{RM}\lambda^2\right) \mathrm{e}^{-2\sigma^2_{\mathrm{RM}}\lambda^4} \label{eq:modelfitU}
    \end{align}
    
    \end{subequations}
    \par These equations can be fit to observed data by creating cubes of the $q$ and $u$ images from the wide band observations. This fitting yields calculated values of $\chi_0$, $\mathrm{RM}$, $\sigma_{\mathrm{RM}}$, and $p_0$ for each pixel.

\begin{figure}
\includegraphics[width=\columnwidth]{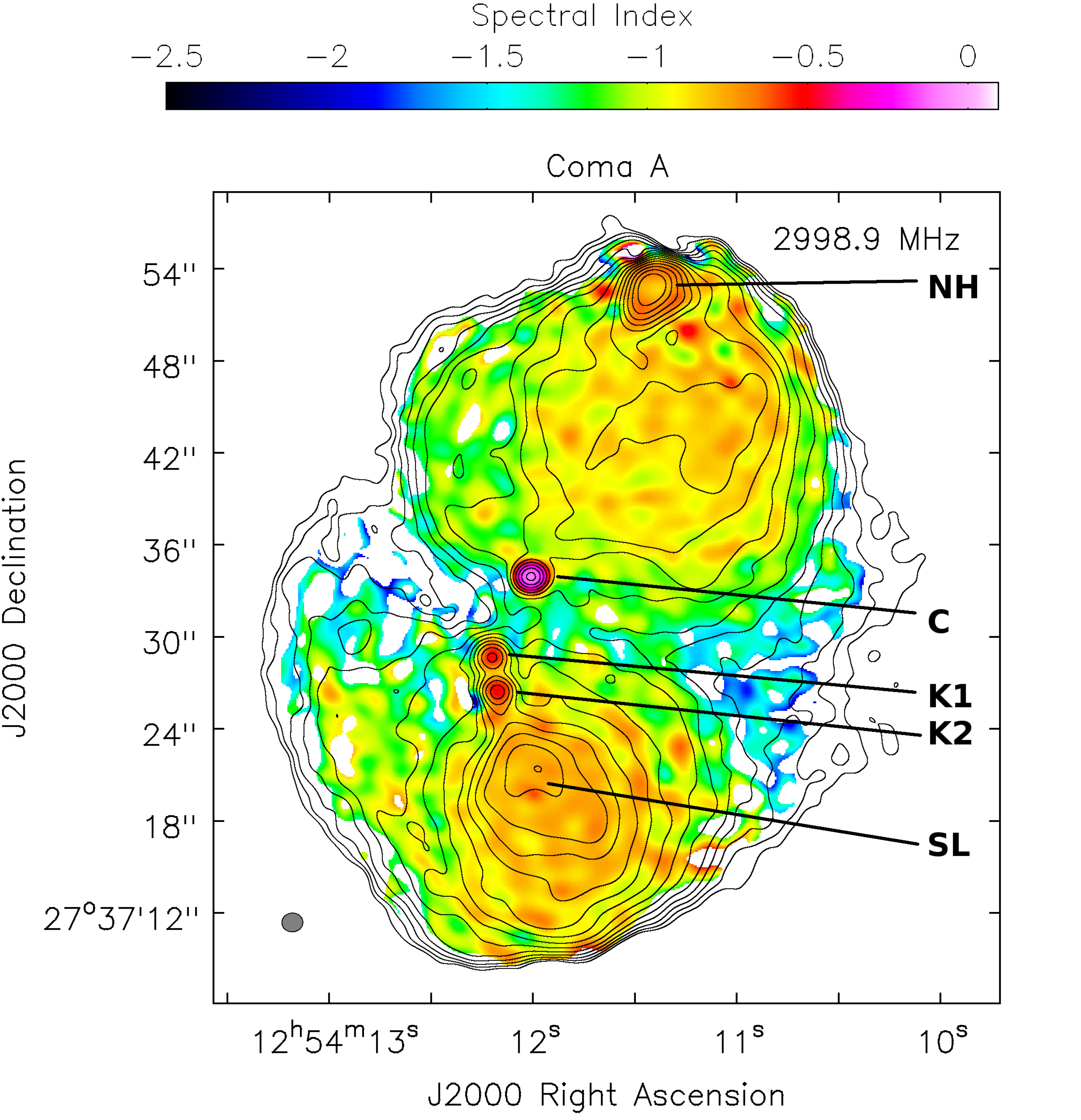}
\caption{The 2--4\,GHz data imaged using a multi-frequency synthesis method with a reference frequency at  3\,GHz in order to observe the most structure with the best possible resolution. The contours are overlaid on the spectral index map used in the imaging process. The convolving beam is 1.37\arcsec $\times$ 1.23\arcsec~ at a position angle of -88.6\degr. The labeled features are: The northern hotspot (\textbf{NH}), core (\textbf{C}), knots (\textbf{K1} and\textbf{ K2}), and the southern lobe (\textbf{SL}). The contours in Stokes $I$ are 0.2\,mJy\,beam$^{-1} \,\times $(1, 1.41, 2, 2.83, 4, 5.66, 8, 11.31, 16, 22.63, 32, 45.25, 64)}\label{fig:Sband_Icont_spix}.

\end{figure}  



\section{Observations and data reduction}

    \subsection{Wide band VLA observations}
    \label{subsec:JVLA_obs}
    
    The  Karl G. Jansky Very Large Array (VLA)  observations in L (994 -- 2006\,MHz) and S (1988 -- 4012\,MHz) bands were taken on the 29 December 2013 and 27 September 2013 respectively, in the VLA B configuration. The flux, phase and bandpass calibrations were done with regular scans of 3C 286, which is close to Coma A, every 20 minutes. A summary can be seen in Table \ref{tab:new_obs_summary}, where $\nu_1$ is the starting frequency of the observing band and $\nu_2$ is the final frequency. The width of each channel in the wideband setup is $\Delta\nu$. The date is the observing date while $t$ is the total integration time on Coma A for that observation.
    
    \begin{table}
    \resizebox{\columnwidth}{!}{\begin{minipage}{\columnwidth}

    \caption{Table outlining the VLA observations of Coma A.}
    \label{tab:new_obs_summary}
    \begin{tabular}{ccccccc}
    \hline 
    Configuration & $\nu_{1}$ & $\nu_{2}$ & $\Delta\nu$  & Date & $t$ \\ 

     & MHz & MHz & MHz &  & hrs\\ 
    \hline \hline
    VLA B & 1988 & 4012 & 2 & 2013 Sep 27 & 1.75$^a$\\ 
 
   VLA B & 994 & 2006 & 1 & 2013 Dec 29 & 1.79\\ 
    \hline
    \end{tabular}\\
    {$^a$ Due to an unknown error in the instrumental polarization calibration, only the final two thirds of these scans had correct polarization calibration applied and were useable for polarization measurements. The Stokes $I$ data are unaffected.}
    \end{minipage}}
    \end{table} 

	The wide band VLA data were flagged, calibrated, and imaged using the Common Astronomy Software Applications package \textsc{casa}. 
	The data were first Hanning smoothed to isolate the RFI interference peaks and remove Gibbs ringing. The data were then flagged for RFI using the standard automated flagging procedures \textit{TFCrop} and \textit{RFlag} in \textsc{casa}. 

     \par To calibrate the instrumental polarization and correct for the polarization angle, a polarized model for 3C 286 was used, as polarization angle and fractional polarization are known for this source. This was done (using \textit{setjy} and \textit{polcal} in \textsc{casa}) because due to scheduling issues the parallactic angle coverage for these observations was not sufficient to solve for the calibrator polarization independently.
Due to an unresolved issue with the polarization leakage calibration, only two thirds of the scans of Coma A for the  2--4\,GHz data had stable amplitudes and phases for polarized correlations. This was indicated by noisy and irregular values in the corrected amplitudes and phases of the scans for 3C 286 adjacent to scans of Coma A. The 2--4\,GHz data that did not display stable polarized amplitudes and phases were not used in the analysis.    

	\par The data were self calibrated over the full spectral range for each dataset using a multi-term multi-frequency synthesis method for imaging. This was done to reduce the noise level in the image, and the wide-band image provided the most accurate model for self calibration.

   \par  For the model fitting, data cubes of  $Q$ and $U$ images were made by imaging the data in discreet frequency bins of 8\,MHz. A Clark clean was used for each separate Stokes parameter. The visibility data were tapered in the imaging process and the resulting images smoothed all to the same beam size of 7\arcsec , which corresponded to the  1\,GHz image. This resulted in data cubes with 126 images along the spectral axis from 1 to 2\,GHz.

    \par For the 2--4\,GHz data, the instrumental polarization calibration was not ideal; the quality of the per-channel instrumental polarization calibration was poor, due to the small range of parallactic angle coverage for the polarization leakage calibrator source. As a result significantly larger frequency bins were used in the imaging process. This was done in an attempt to reduce the effects of the Stokes $Q$ and $U$ values varying over each spectral window. The visibility data were tapered and images smoothed so that all resulting images had an identical 4.1\arcsec~\, circular beam

    Cubes of  Stokes $Q$ and $U$ images were made with frequency bins of 128\,MHz from 2 to 4\,GHz. The resulting cubes have 16 evenly spaced images along the frequency axis.

    \subsection{Archival VLA observations}
    
    To examine the behaviour at higher frequencies, and confirm the observed polarization structure at lower frequencies archival observations  at 4.9 and 15\,GHz were used. The archival data are also at a higher resolution than the more recent wide-band observations. The raw data were downloaded from the VLA archive and calibrated again for this publication. These observations were useful in localising exact regions where depolarization occurs. A summary of the archival data analysed is given in Table~\ref{tab:old_obs_summary}, where $\nu$ is the observing frequency. The width of each channel in the wideband setup is $\Delta\nu$. The date is the observing date while $t$ is the total integration time on Coma A for that observation.
    \begin{table}
    \resizebox{\columnwidth}{!}{\begin{minipage}{\columnwidth}

    \caption{Table outlining the archival VLA observations of Coma A. }
    \label{tab:old_obs_summary}
    \begin{tabular}{ccccccc}

    \hline 
    Project & $\nu$ & $\Delta\nu$ & Configuration & Date & $t$ \\ 

     & MHz & MHz &  &  & hrs\\ 
    \hline \hline
    
    VANB$^a$ & 4885 & 50 & VLA B & 1981 Jun 01 & 4.1\\ 
 
    AB348 & 14940 & 100$^{b}$ & VLA C & 1985 Aug 26 & 6.2\\ 
    \hline
    \end{tabular}\\\\
    {$^a$ Data previously published in \cite{VanBreugel1985}\\$^{b}$ Data consist of two frequencies 50\,MHz apart each with  50\,MHz bandwidth. Central frequency is given.}
    \end{minipage}}
    \\
    \\
    \end{table}

In addition to confirming the results of the VLA observations, the archival VLA data were used to investigate the polarization behaviour at higher frequencies and resolutions, where higher fractional polarization was expected. These were calibrated using the Astronomical Image Processing System (\textsc{aips}) 
software package and the resulting visibility data were exported to \textsc{casa} for imaging using a multiscale CLEAN. Images of the polarized intensity were made with the standard polarisation bias correction applied in CASA (e.g. VLA memo 161, \cite{Leahy1989} ). The fractional polarization images were made using only Stokes $I$ as a blanking factor, in this case a blanking limit of 1\,mJy\,beam$^{-1}$ in Stokes $I$ was used. This was important for mapping the heavily depolarized regions that are present even at high frequencies (4.9 and 15\,GHz).

	\subsection{H\,$\alpha$ observations}
   
   The deep H\,$\alpha$ image used in this paper is taken from \cite{Tadhunter2000}, which was observed with the William Herschel Telescope in January 1998. The image is the result of a deep $2\times 900$ second integration and has been continuum subracted to show the pure H\,$\alpha$ emission. Further details about the observations and the calibration of this data can be found in \cite{Tadhunter2000}.

    \section{QU model fitting procedure}

To analyse the polarization and Faraday rotation properties of the lobes of Coma A, we fit the data with an external Faraday dispersion model, as described in Section \ref{subsec:RM_QU_theory}. However, to initially identify regions with significant polarization the cubes of Stokes $Q$ and $U$ images were first processed using RM synthesis \citep{Brentjens2005}. This was done using the python implementation of \textit{pyrmsynth}\footnote{ \url{http://mrbell.github.com/pyrmsynth} }.
    The resulting total polarized flux image from the RM synthesis procedure was used to make a mask for the $Q$ and $U$ cubes. Only regions where the polarized flux in this image was greater than three times the noise level were included in the $Q$ and $U$ cubes. This was done as initial inspections of model fits showed that frequency dependant instrumental polarization was contaminating the data in regions of low signal to noise. It must also be noted that due to this instrumental polarization effect, the 1--2\,GHz and 2--4\,GHz datasets were analysed and modeled independently.
    
    \par Before the model fitting, the $Q$ and $U$ cubes were converted to fractional $Q$ and $U$ cubes (referred to as $q$ and $u$). This was done using model Stokes $I$ images constructed from a spectral index map, which was made from the multi-frequency synthesis imaging, and an image at the central frequency for that band. This was done to avoid regions with negative Stokes $I$ arising from dynamic range issues due to the brightness of Coma A and the narrow bands used in making the cubes. This also limited the propagation of noise fluctuations in the Stokes $I$ images to the $q$ and $u$ data. 
    
  \par  The cubes were fit according to the model in Equations \ref{eq:modelfitQ} and \ref{eq:modelfitU}. 
A single component RM model was used with a depolarization term from an external screen, which is what was expected from the H\,$\alpha$ measurements and previous publications indicating a gas shell surrounding the lobes. 
    The fitting was done for every unmasked pixel in the cubes using an ``asexual genetic algorithm'' \citep{Canto2009}, which is very efficient in converging to the global best-fit. The best-fitting model values were quantified by minimizing the chi-squared function for each fit in each pixel. The results of this fitting were values of $\chi_0$, $\mathrm{RM}$, $\sigma_{\mathrm{RM}}$, and $p_0$ for each unmasked pixel. The pixels from the resulting maps were also blanked according to the error in the model fit.  For this the error in the $\sigma_\mathrm{RM}$  was a good discriminator between acceptable model fits and those contaminated by any remaining residual leakage errors in the low polarization regions.
    
\begin{figure*}
\includegraphics[width=2\columnwidth]{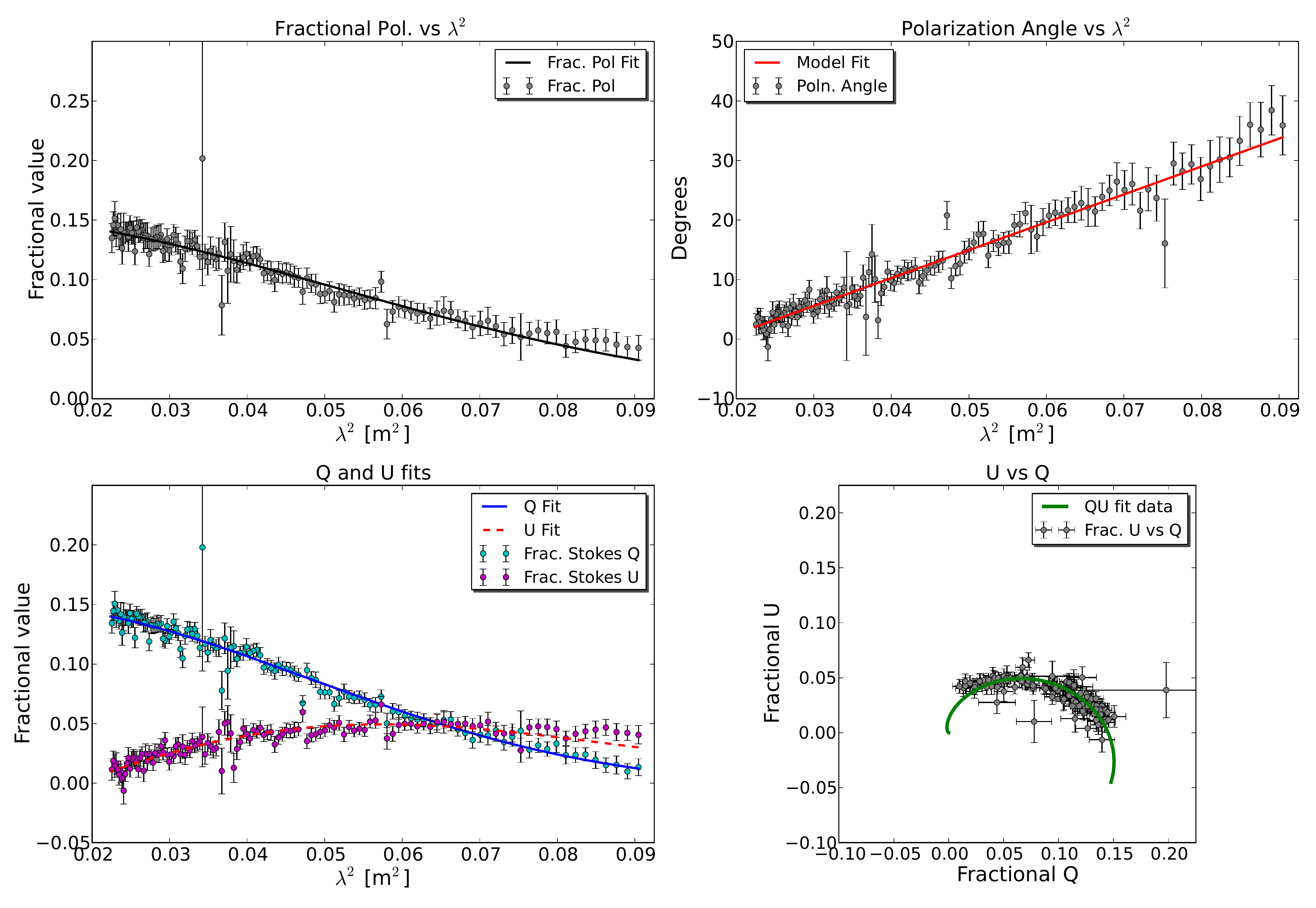}
\caption{Model fit for the 1--2\,GHz data. This is in the upper north western region of the southern lobe, where the polarized flux is the highest. The bottom left panel shows the individual fractional Stokes $Q$ and $U$ fits. The upper left panel shows how the fractional polarization changes with wavelength squared ($\lambda^2$). This is made using $p = \sqrt{q^2+u^2}$ from the fit data in the panel below it. Similarly for the top right panel the actual polarization angle data is plotted. The line passing through is calculated from the fit data as $\chi = 0.5\arctan\left(u/q\right)$, not a linear fit. The bottom right panel simply plots the fractional Stokes $Q$ and $U$ against each other with model fitted data overlaid, the fitted line has been extended in $\lambda^2$ to show the fit spiralling toward the origin due to the depolarization. From this fitted region the fit values are: $p_0 = 0.15 \pm 0.001$, $\mathrm{RM} = 3.9 \pm 0.1 \,\rm{rad\,m}^{-2}$, $\sigma_\mathrm{RM} = 9.8 \pm 0.1\, \rm{rad\,m}^{-2}$, $\chi_0 =  -8.5 \pm 0.2\degr$.}\label{fig:model_fit_L}
\end{figure*}  

\begin{figure*}
\includegraphics[width=2\columnwidth ]{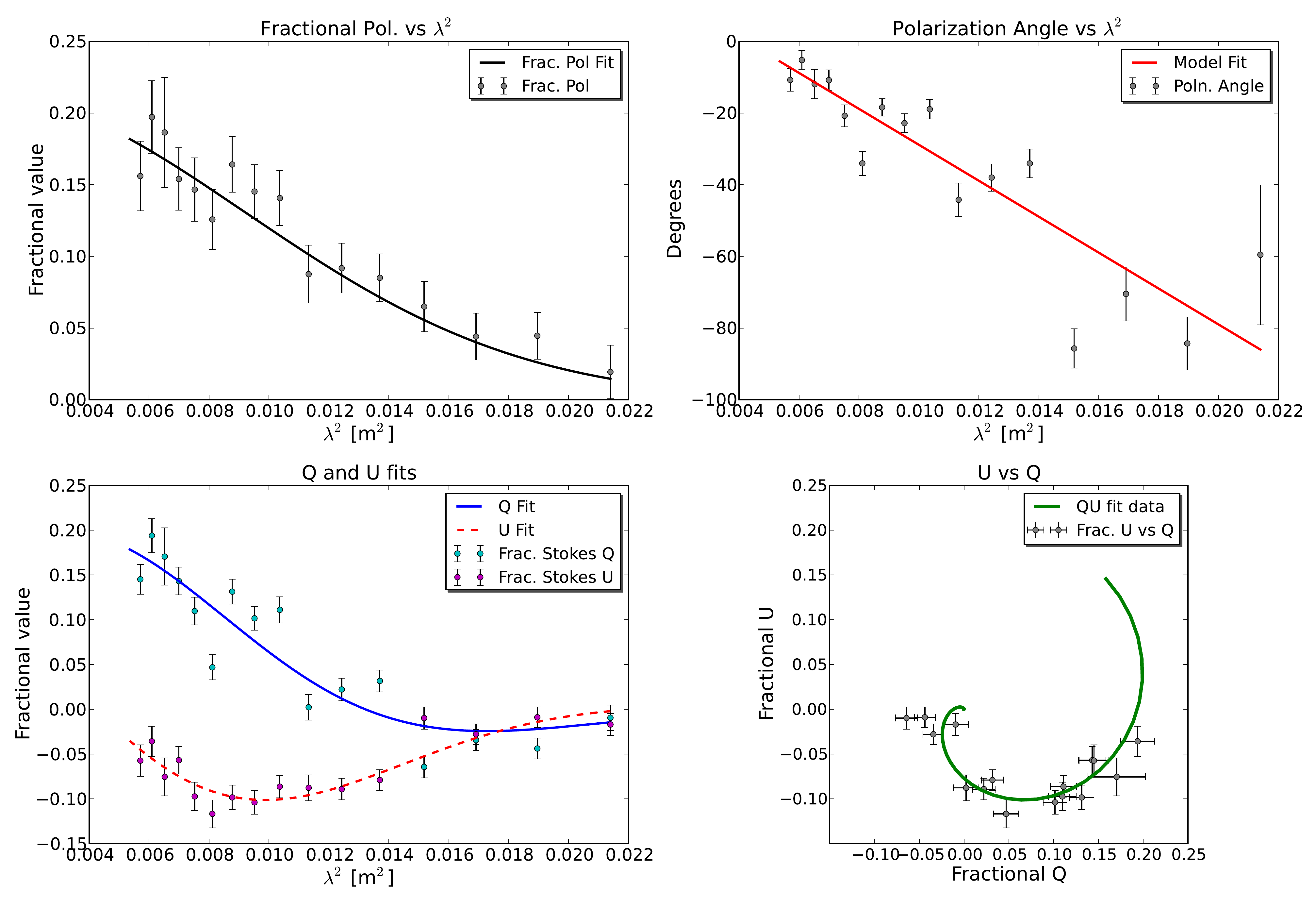}
\caption{Model fit data for the 2--4\,GHz data. This is for a region just to the south west of the H\,$\alpha$ filament in the southern lobe. Note the strong depolarization. The graphs panels are displayed in an identical manner as in Figure~\ref{fig:model_fit_L}. From this fitted region the fit values are: $p_0 = 0.22 \pm 0.03$, $\mathrm{RM} = -88 \pm 7\,\rm{rad\,m}^{-2}$, $\sigma_\mathrm{RM} = 54 \pm 7\,\rm{rad\,m}^{-2}$, $\chi_0 =  -69 \pm 4\degr$.}\label{fig:model_fit_S}
\end{figure*}

\begin{figure*}
\includegraphics[width=\columnwidth]{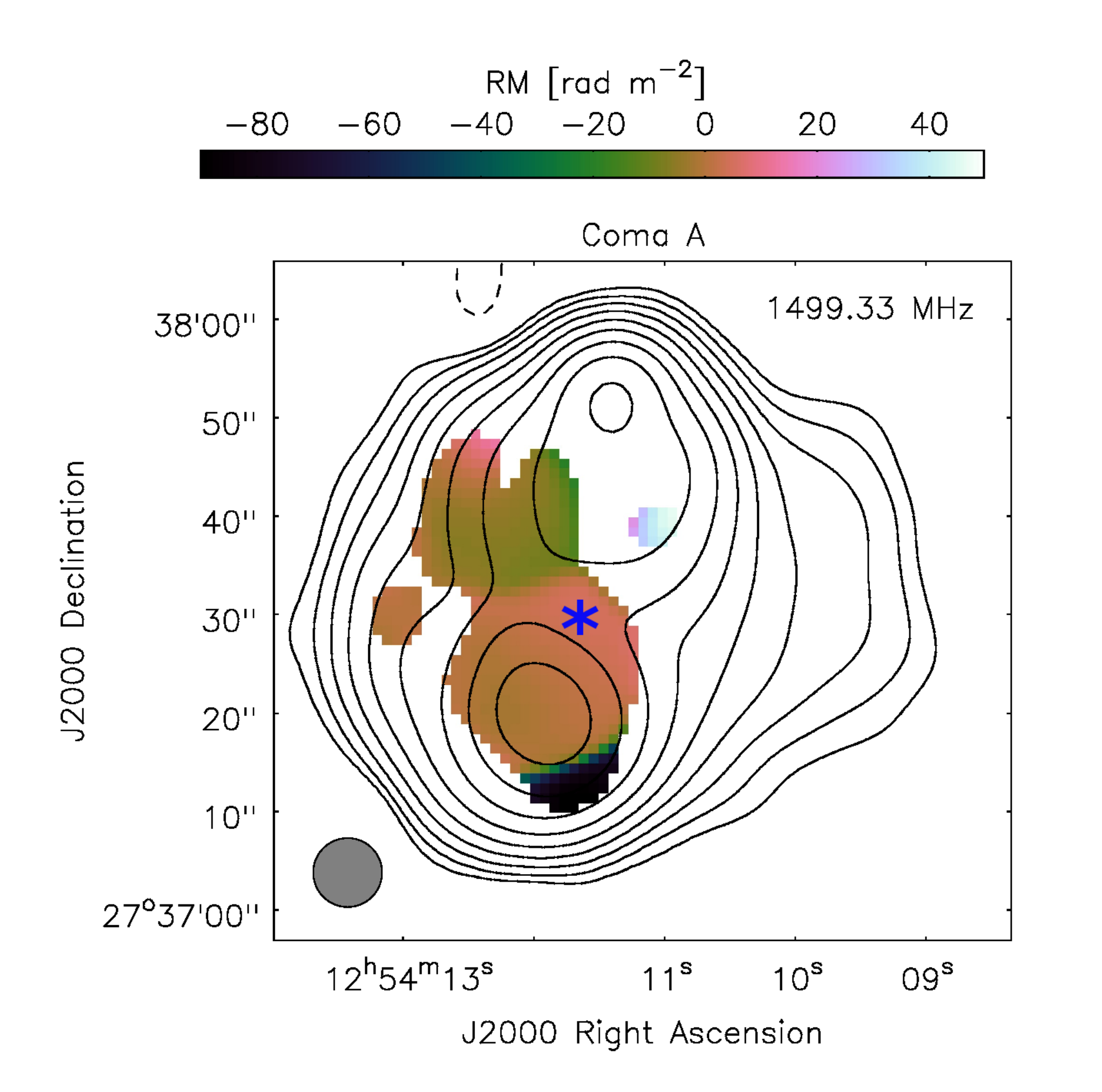}
\includegraphics[width=\columnwidth]{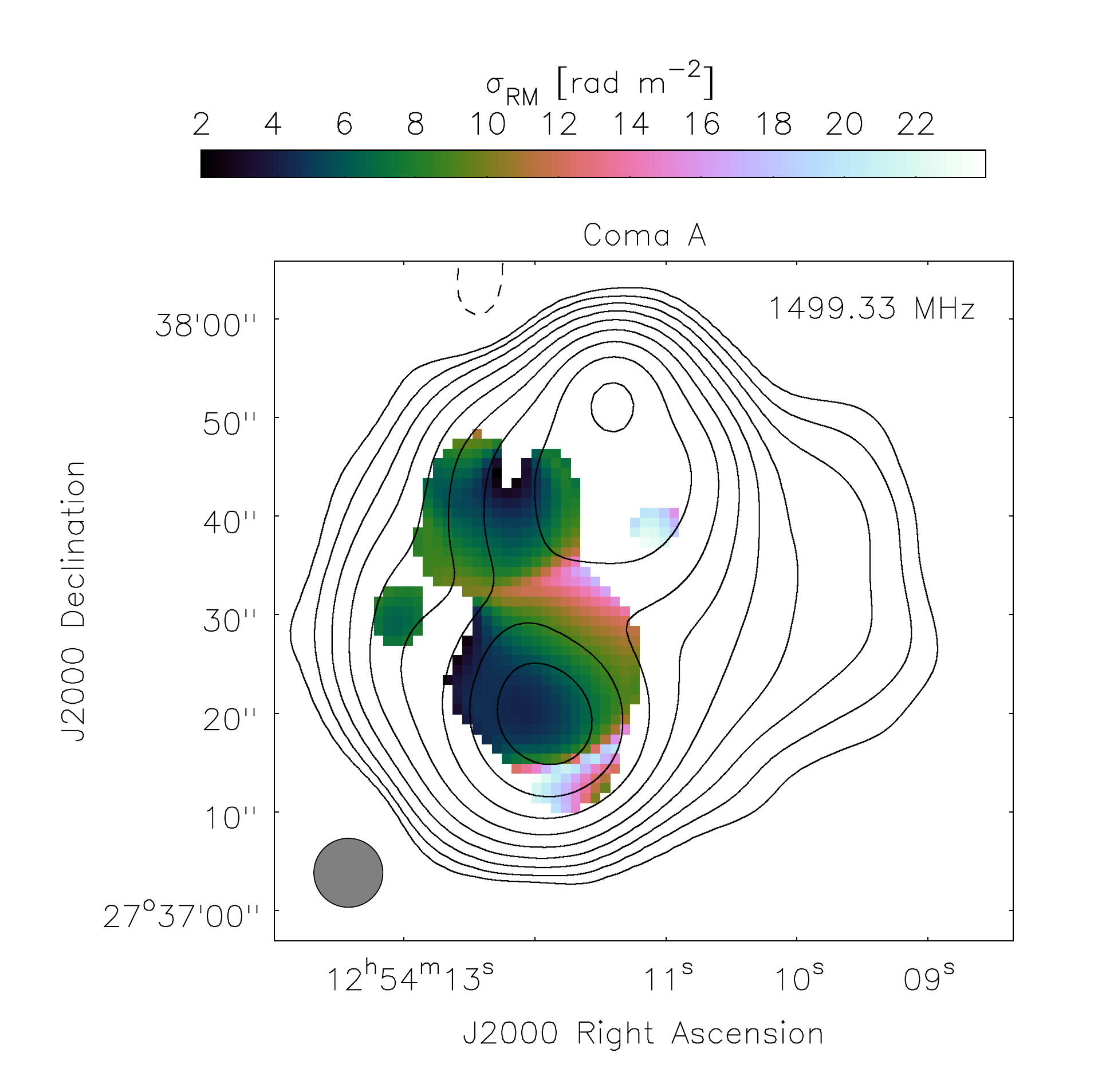}
\caption{These are the RM (left) and $\sigma_\mathrm{RM}$ (right) images produced from the model fitting of the 1--2\,GHz data. Due to the low resolution used, images of the fractional polarization at zero wavelength ($p_0$) and intrinsic polarization angle ($\chi_0$) are not included. Note that the blue~`$\mathbf{\ast}$' is the location for which the data in Figure~\ref{fig:model_fit_L} are plotted. For the Stokes $I$ contours, a 1--2\,GHz image made using a multi-frequency synthesis technique is used. The uv data were  tapered and the image smoothed to the same 7\arcsec~ resolution as the images used in the model fit. The central frequency is given in the top right corner of the images. The contours are given at the levels of 1\,mJy\,beam$^{-1}\,\times$ (1, 2, 4, 8, 16, 32, 64, 128, 256)}
\label{fig:Lband_fit_images}
\end{figure*}

\begin{figure*}
\includegraphics[width=2\columnwidth]{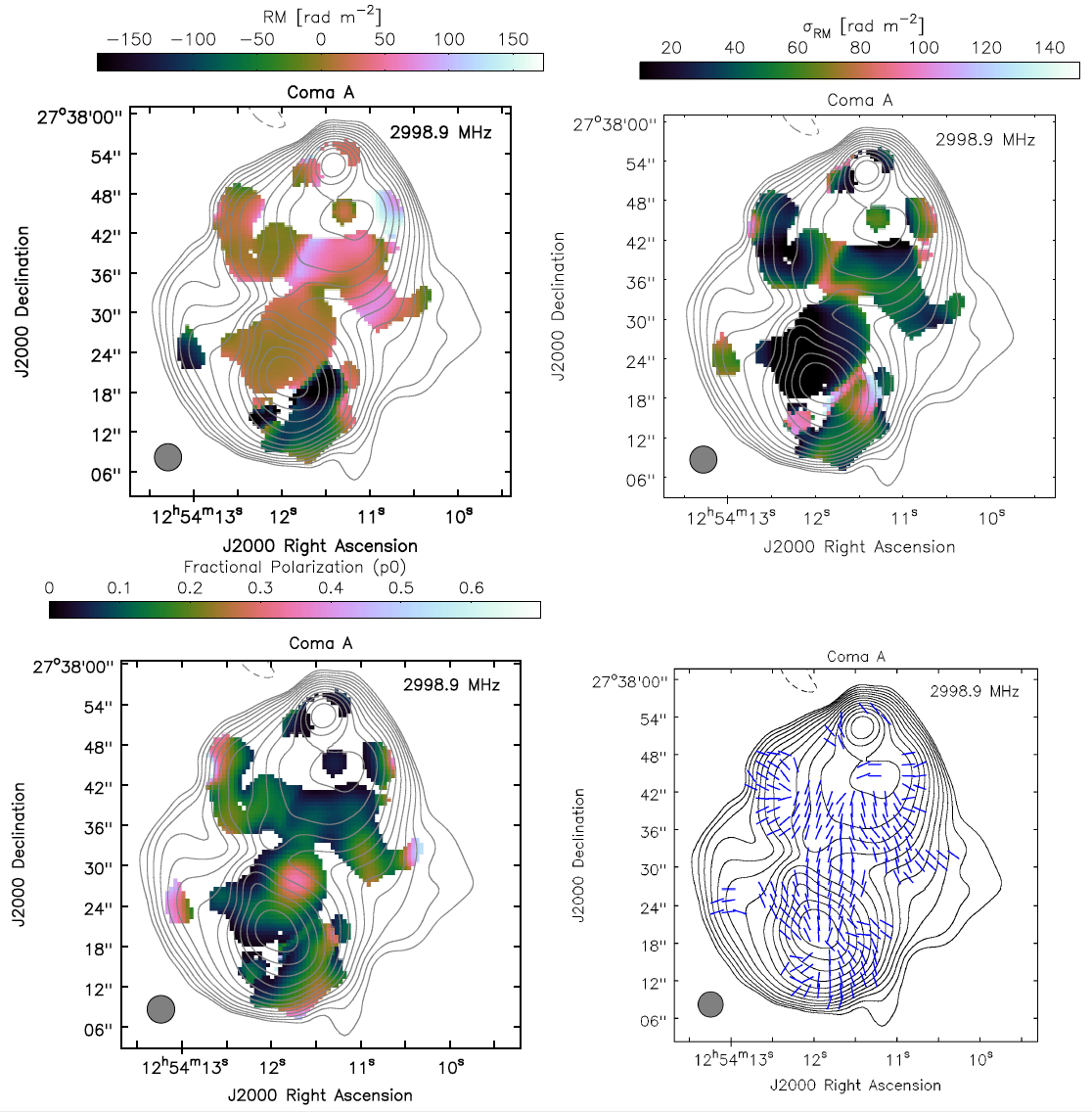}
\caption{These are the RM (top left) and $\sigma_\mathrm{RM}$ (top right) images produced from the model fitting procedure for the 2--4\,GHz data. The images  of the fractional polarization at zero wavelength ($p_0$) and intrinsic polarization angle ($\chi_0$) are given on the bottom left and bottom right respectively. The images were blanked according to the errors in $\sigma_\mathrm{RM}$ as this parameter had the highest errors and was a good indicator of the general quality of a fit for a depolarization model. The Stokes $I$ contours are from a 2--4\,GHz image made using a multi-frequency synthesis method using a central frequency of 3\,GHz. The image was tapered and smoothed to the same 4.1\arcsec~ resolution as images used in model fit.  The contours are given at the levels 1.4\,mJy\,beam$^{-1}\,\times$ (-1, 1, 1.41, 2, 2.83, 4, 5.66, 8, 11.31, 16, 22.63, 32, 45.25, 64) with the negative contour dashed. }
\label{fig:Sband_fit_images}
\end{figure*} 



\section{Results}

    \subsection{Radio structure of Coma A}
    \label{subsec:structofComaA}
    In Figure~\ref{fig:Sband_Icont_spix} we present a Stokes $I$ contour map of Coma A overlaid on the spectral index map. This image was made from the  2--4\,GHz data and imaged using the multi-frequency synthesis method centred on 3\,GHz using three Taylor terms in the \textsc{casa} task \textit{tclean}. The spectral index map, made using only the 2--4\,GHz data was cross checked with an  inter--band spectral index map made using the central frequencies from both the 1--2\,GHz and 2--4\,GHz data (at a lower resolution of 4\arcsec), finding consistent values with the higher resolution 2--4 GHz spectral index map. Additionally the values for spectral index calculated for Coma A in \cite{VanBreugel1985} using 1.4 and 4.9\,GHz data agree with Figure \ref{fig:Sband_Icont_spix} also. The northern hotspot, the knots K1 and K2, which were first identified and named in \cite{Bridle1981}, and the southern lobe have been labelled. The hotspot to the north is typical for an FRII galaxy showing where the jet is impacting with the intergalactic medium. The knot K1 shows the location where the jet is colliding with a gas cloud, which can be identified in optical images (see right hand panel of Figure 8 in \cite{Worrall2016}). The K2 knot and diffuse lobe are due to the deceleration of the jet after the deflection of $\sim 30\degr$ from the radio axis on the plane of the sky. The core has a relatively flat spectrum of $\alpha\simeq -0.1$, while outer features all have $\alpha \simeq -0.7$ or less, showing a steeper spectrum typical of optically thin synchrotron emission and older regions of the jet emission. \\
    
\subsection{QU model fitting}
    \label{subsec:resultsqufit}
    The model fitting showed that a single component RM and external depolarizing random screen provided an excellent fit to the data (Figures~\ref{fig:model_fit_L} and \ref{fig:model_fit_S}). The images of the calculated fitting parameters, RM, $\sigma_\mathrm{RM}$, $p_0$, and $\chi_0$ can be seen in Figures~\ref{fig:Lband_fit_images} and \ref{fig:Sband_fit_images}.
 \subsubsection{Northern lobe}
    
\begin{figure}
	\centering
	\includegraphics[width=0.9\columnwidth]{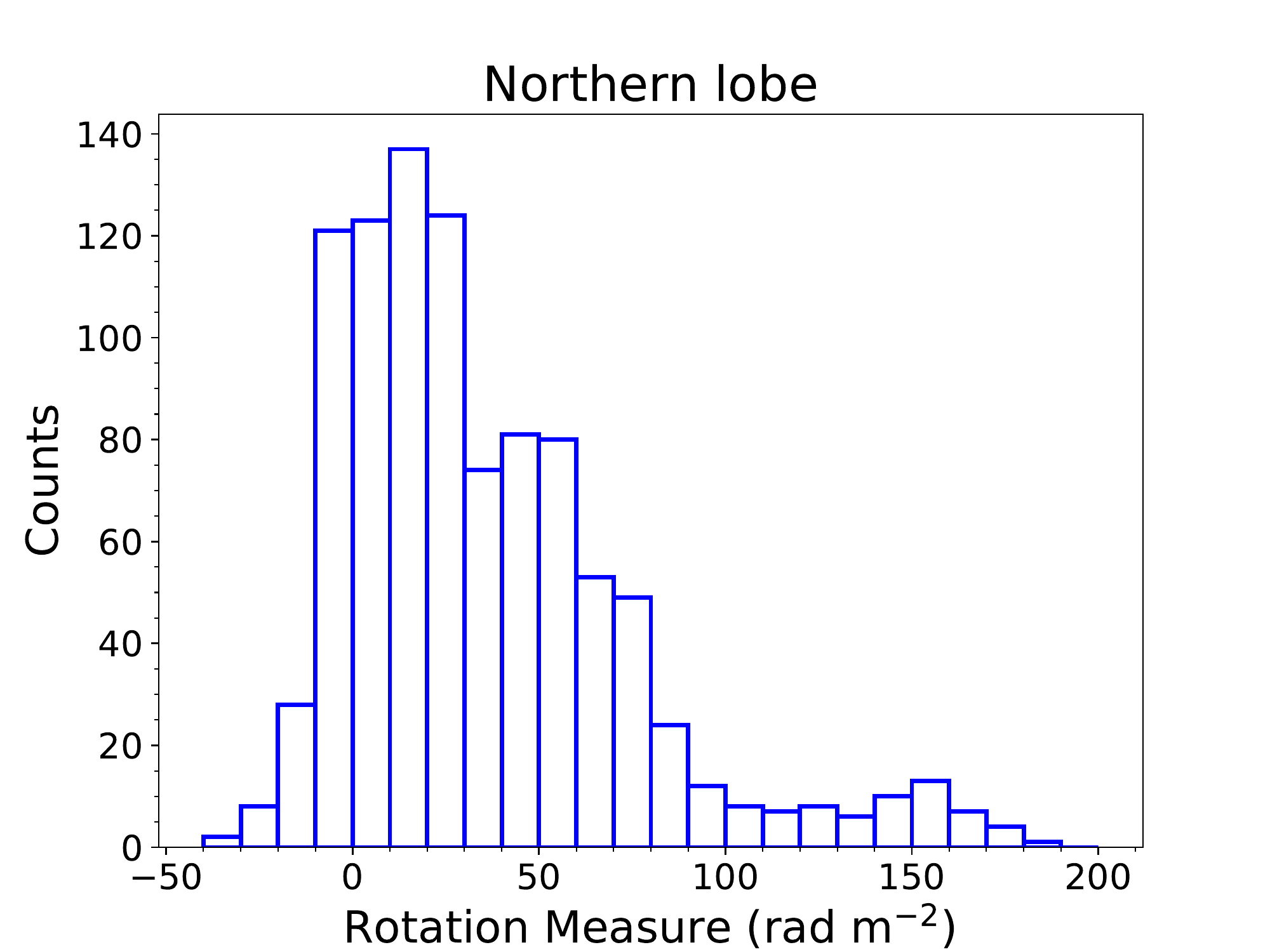}
	\includegraphics[width=0.9\columnwidth]{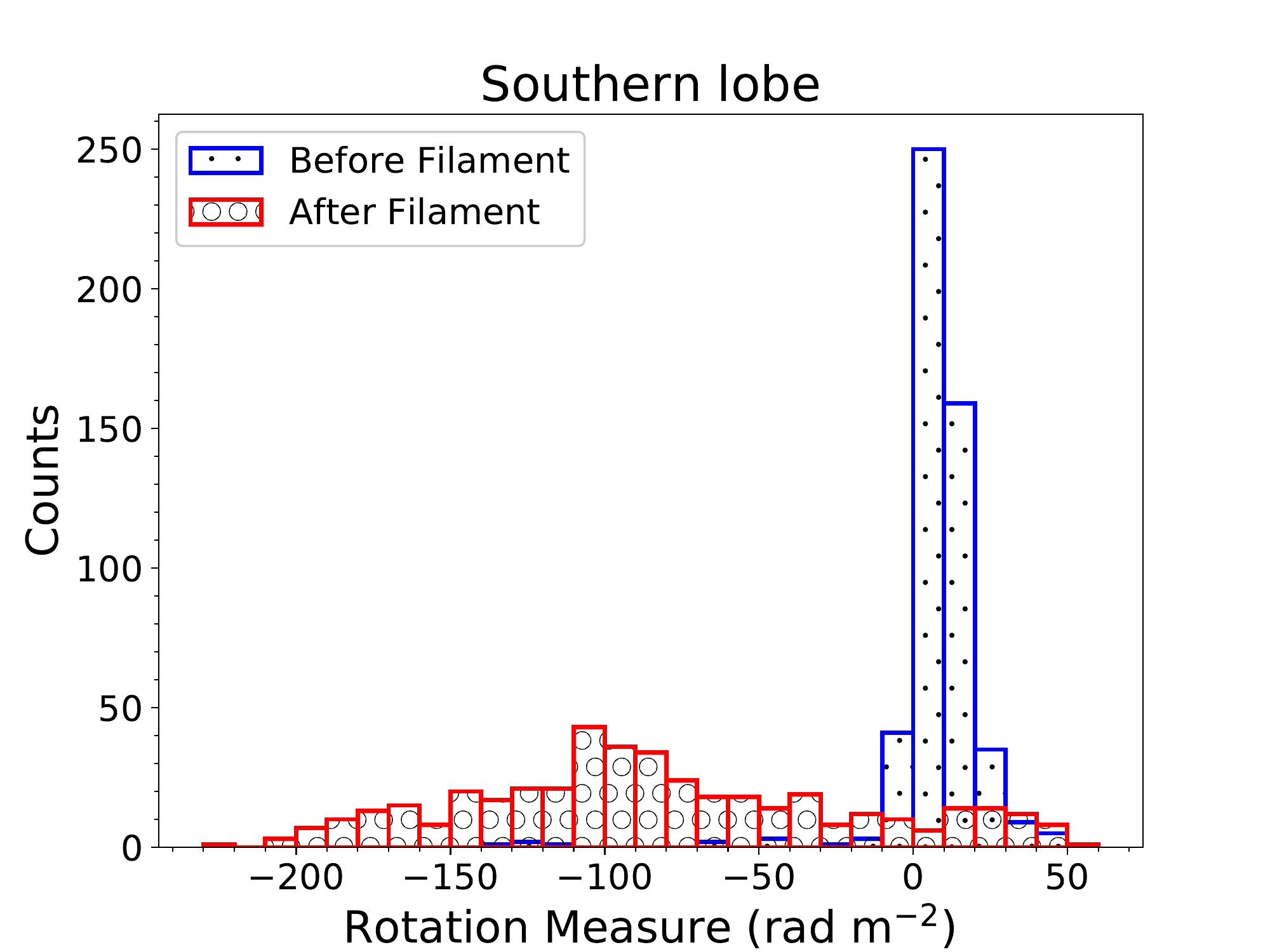}
	\caption{Histograms of the RM distribution in the lobes of Coma A taken from the RM image made for the S-Band model fit results. The top panel is the histogram of the RM distribution across the Northern lobe. The bottom panel shows the RM distribution in the Southern lobe, the blue histogram represents the part of the jet before the depolarizing H\,$\alpha$ filament; the red histogram represents the RM values after the filament and including the RM feature to the east.}
	\label{fig:histograms}
\end{figure}

    The northern hotspot did not have detectable polarization ($\leq 0.4\%$) in the 1--2\,GHz range making the model fits poor. This is most likely due to depolarization from large RM variations in addition to intrinsic differences in polarization angle within the 7\arcsec~ beam across the complex hotspot structure. The fractional polarization for the hotspot was also low in the 2--4\,GHz observation where only some polarization is seen to the immediate east and west of the hotspot (ranging from $\sim 2\%$ to $4\%$). The centre of the hotspot has been blanked due to high error in the fits and low polarization. To confirm this, the archival dataset at 4.9\,GHz (Figure~\ref{fig:archival_fpolHa}(a) ) was used as a reference. The polarized intensity at the exact hotpot is too faint to detect ($<1$\%), only to the North and South of the hotspot is polarization reliably measured with values ranging between 2\% and 9\%, this confirms the results for the wideband 2--4\,GHz observations. The 15\,GHz observations at a similar resolution shows some polarization (Figure~\ref{fig:archival_fpolHa}(d)) for the hotspot at around 15\% however this frequency corresponds to a $\lambda^2$ value of $4\times 10^{-4}\,\rm{m}^2$ which is close to zero wavelength. This indicates there is strong depolarization in the hotspot region. Note also that the lower 4\arcsec~ resolution used in the wide band 2--4\,GHz observations may also lead to additional depolarization.
    \par The Northern lobe as a whole shows a patchy polarization distribution, presumably due to Faraday depolarization from the H\,$\alpha$ emitting gas, as the polarized emission is strongly anti-correlated with the H\,$\alpha$ emission (Figure~\ref{fig:archival_fpolHa}(b) and (d)). The RM distribution is shown in the histogram in the top panel of Figure~\ref{fig:histograms} with a mean (median)  RM of 34.9~(24.9)\,rad\,m$^{-2}$ and standard deviation (median absolute deviation)  of 39~(22.6)\,rad\,m$^{-2}$. The high depolarization in the northern hotspot and lobe as a whole can be interpreted as an example of the Laing--Garrington effect \citep{Garrington1988,Laing1988}, which implies that the northern jet is receding relative to the observer and the southern, more strongly polarized jet is approaching. The fat projected shape of the lobes is also consistent with an orientation significantly inclined to the plane of the sky. One caveat to this is that the southern jet, having been deflected by the collision in K1, has altered its orientation significantly relative to the inner radio axis.

\subsubsection{Southern lobe}
    The polarization data for the southern lobe agrees very well with the model in the 1--2\,GHz range, as can be seen, for example, in Figure~\ref{fig:model_fit_L} (The data corresponds to the location indicated by a the asterisk in Figure~\ref{fig:Lband_fit_images}).  At this pixel location,  $p_0 = 0.15 \pm 0.001$, $\mathrm{RM} = 3.9 \pm 0.1 \,\rm{rad\,m}^{-2}$, $\sigma_\mathrm{RM} = 9.8 \pm 0.1\, \rm{rad\,m}^{-2}$, and $\chi_0 =  -8.5 \pm 0.2\degr$.  The model fit parameters from the  2--4\,GHz data, at the same location, are $p_0 =0.33 \pm 0.03$,  $\mathrm{RM} = 12 \pm 4\,\rm{rad\,m}^{-2}$, $\sigma_\mathrm{RM} = 25.6 \pm 6\,\rm{rad\,m}^{-2}$ and $\chi_0 =  -14 \pm 3\degr$. 
    This shows the value of the higher resolution data at 2--4\,GHz (4\arcsec~beamsize) which more robustly probes the underlying Faraday rotation variations across the lobes. In particular, the higher value of $p_0$ indicates that we have resolved most of the RM structure in this region, as the fractional polarization at 15\,GHz in this region is $\sim$0.35 (Figure~\ref{fig:archival_fpolHa}(c)). Considering that the model of a Faraday depolarizing screen fits well to the data, the difference in the RM and $\sigma_\mathrm{RM}$values between 1--2\,GHz and 2--4\,GHz can be ascribed to resolution dependent effects for an inhomogeneous Faraday screen produced by the patchy distribution of H\,$\alpha$ across the lobes.     
            
	\par Looking at the fractional polarization images for the higher resolution archival observations at 4.9\,GHz and 15\,GHz, there are two regions of appreciably low fractional polarization. Firstly there is a region of low polarization along the centre of the jet in the region after the knot K2. The fractional polarization at 15\,GHz varies from $\sim14\%$ in the centre to over $30\%$ on either side of this region. This may be due to entrainment of external medium into the jet after deceleration following the collision with K1 \citep{VanBreugel1985} or wavelength-independent depolarisation due to the jet magnetic field orientation being roughly orthogonal to the magnetic field orientation in the surrounding lobe. However a toroidal or helical field in the jet giving high fractional polarization at the edges may be another explanation for this feature, which is discussed in \cite{knuettel2017}.

\begin{figure*}
\includegraphics[width=2\columnwidth]{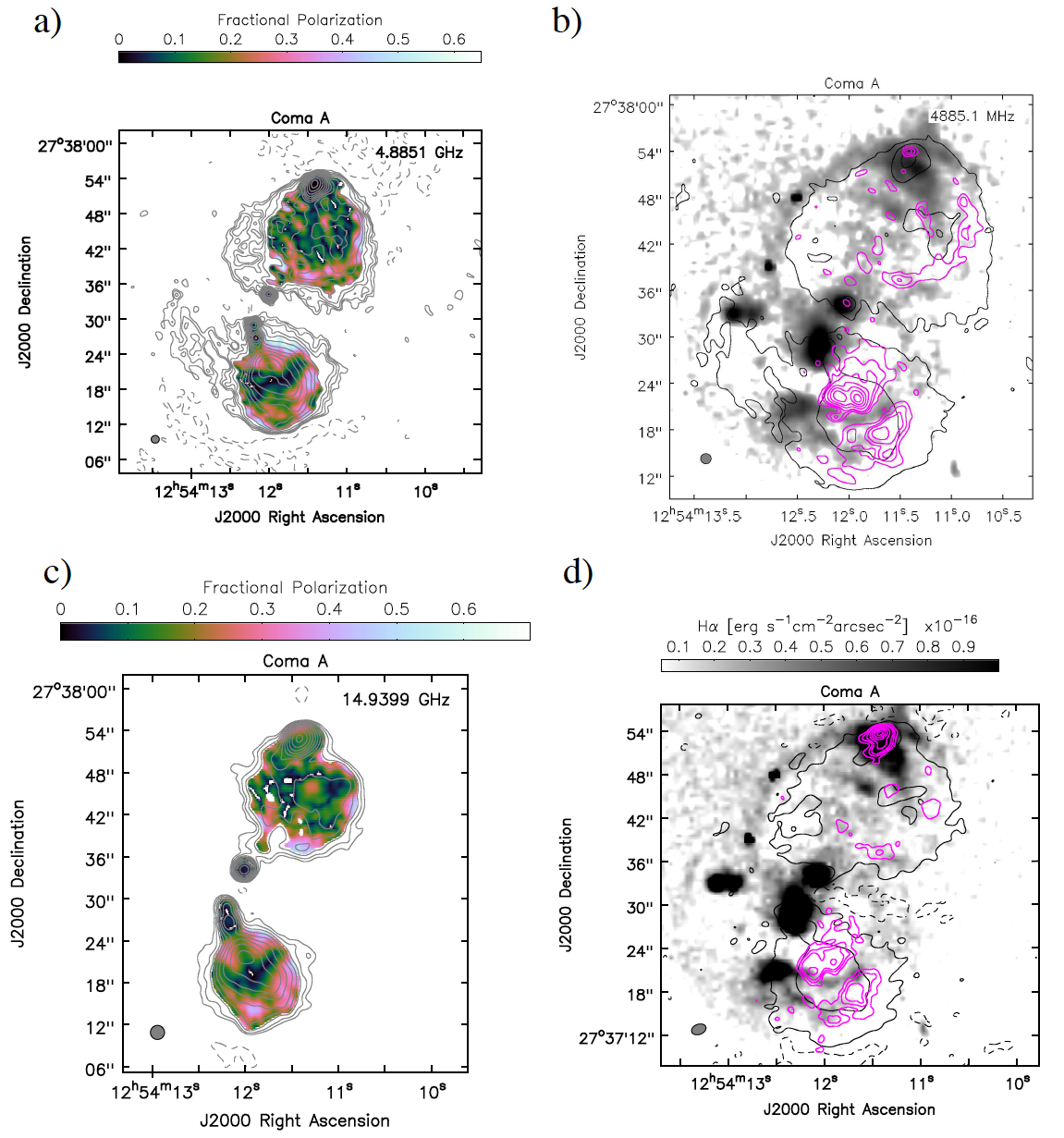}
\caption{\textbf{(a)} 1.4\arcsec~resolution 4.9\,GHz image of Coma A from the VLA archival data. The colour image is the de-biased fractional polarization  where the corresponding Stokes $I$ intensity is greater than 1\,mJy\,beam$^{-1}$at 4.9\,GHz  and the contours are the total Stokes $I$ emission at that same frequency. The contours are given at the levels of 0.25\,mJy\,beam$^{-1}\,\times$ (-1, 1, 1.41, 2, 2.83, 4, 5.66, 8, 11.31, 16, 22.63, 32, 45.25, 64, 90.51) with the negative contour dashed.  \textbf{(b)} Image of Coma A showing polarized flux density in Jy\,beam$^{-1}$ as purple contours with levels at 0.45\,mJy\,beam$^{-1}$ $\times$ (1, 1.5, 2, 3, 4, 5, 6). Stokes $I$ contours are given in black at 27.5\,mJy\,beam$^{-1}$ $\times$ (0.008, 0.2, 0.6) which corresponds to 0.8, 20 and 60 percent of the peak value. The deep H\,$\alpha$ image is given in greyscale. \textbf{(c)}~2\arcsec~resolution 15\,GHz image of Coma A from the VLA archival data. The colour image is the debiased fractional polarization at 15\,GHz displaying only values where the corresponding Stokes $I$ value is greater than 1\,mJy\,beam$^{-1}$. The contours are the total Stokes $I$ emission at that same frequency.  The contours are given at the levels of 0.45\,mJy\,beam$^{-1}\,\times$ (-1, 1, 1.41, 2, 2.83, 4, 5.66, 8, 11.31, 16, 22.63, 32) with the negative contour dashed. \textbf{(d)} Image of Coma A showing polarized flux density at 15\,GHz in Jy\,beam$^{-1}$ as purple contours with levels at 0.4\,mJy\,beam$^{-1}$ $\times$ (1, 1.5, 2, 3, 4, 5, 6, 7). Stokes $I$ contours at the same frequency are given in black at 0.2\,mJy\,beam$^{-1}$ $\times$ (-1, 1 , 8, 32) with the negative contour dashed. The deep H\,$\alpha$ image is given in greyscale.}
\label{fig:archival_fpolHa}	
\label{fig:5ghz_Ha_pol}
\label{fig:Uband_fpol}	
\label{fig:15ghz_Ha_pol}
\label{fig:archival_fpolHa}	
\end{figure*}

    The second feature is a `trough' in polarized flux in a band transverse to the southern lobe,  visible in both the 4.9\,GHz and 15\,GHz fractional polarization images ( Figure~\ref{fig:archival_fpolHa}(a) and ~\ref{fig:archival_fpolHa}(c)).  At 15\,GHz, the fractional polarization in the `trough' is estimated to be $\leq3\%$. What is interesting about this feature is how the lack of polarized emission in this `trough' correlates with a large filament of H\,$\alpha$ emission. Figures~\ref{fig:archival_fpolHa}(b) and \ref{fig:archival_fpolHa}(d) and  show images of the polarized flux and H\,$\alpha$ emission for both frequencies. This indicates that a random magnetic field component coupled with a high electron density within a beamwidth in this `trough' is causing very significant depolarization, even at 15\,GHz. An estimate of the magnetic field and depolarization in this region is made in Section \ref{subsec:bfield_est_theory}.          
    \par The maps of the various parameters from the model fitting in  Figures~\ref{fig:Lband_fit_images} and \ref{fig:Sband_fit_images} do not have a high enough resolution to show the `trough', however, the parameters change drastically  downstream in the jet after this feature. This can be seen in the histogram plots of the RMs for these regions (see bottom panel of Figure~\ref{fig:histograms}). Upstream from the filament the mean (median) RM is 8.4~(7.7)\,rad\,m$^{-2}$ with standard deviation (median absolute deviation from the median) of 10~(4.9)\,rad\,m$^{-2}$, whereas after the filament, the  RM has a mean (median) value  of  -83.3~(-90.3)\,rad\,m$^{-2}$ and much higher standard deviation (median absolute deviation) of 75.4~(40.3)\,rad\,m$^{-2}$. The region after this filament also has a larger amount of Faraday depolarization, as can be seen from the data and model fit in Figure~\ref{fig:model_fit_S}. The large depolarization of $\sigma_\mathrm{RM} =54\pm7 \rm{rad\,m}^{-2}$, is typical for this region and can explain why only a small part of this region is detected in the 1--2\,GHz data (Figure~\ref{fig:Lband_fit_images}). The systematically higher values of $\sigma_\mathrm{RM}$ after the filament can also be seen in the map of $\sigma_\mathrm{RM}$ values in Figure~\ref{fig:Sband_fit_images}.

 	\par This drastic change in the RM downstream from the jet in the southern lobe is an indicator that the jet/lobe has undergone a significant change in morphology or has interacted with external media. The presence of the H\,$\alpha$ filament suggests the latter, especially as this behaviour is not mirrored in the northern lobe.  The evidence provided by \cite{Morganti2002} that the lobe is expanding into a disk of neutral gas and ionising it, creating the filaments observed, is a viable explanation for the depolarizing `trough' feature. However, a potential complication to any models where there is a giant gas disk perpendicular to the radio jet axis (e.g. \cite{Gopal-Krishna2000a}), is the likely large inclination of the radio jet axis with respect to the line of sight (as evidenced by the fat-double shape of the lobes, and the Laing-Garrington effect). This would imply that any H\,$\alpha$ filament related to this disk material is most likely behind the southern lobe, unless the southern jet has a large deflection back into some of this disk material. Further observations of the complex gas dynamics in this system are warranted.

    \subsection{Estimation of magnetic field strengths}
    \label{subsec:bfield_est_theory}
    The depolarization by the H\,$\alpha$ gas makes it challenging to detect the polarized emission. However, in some regions where there are reliable RM, $\sigma_{\mathrm{RM}}$, and H\,$\alpha$ values, an estimate of the magnetic field strength within the ionised gas shell can be made. The strong anti--correllation between the H\,$\alpha$ emission and the polarization (Figure \ref{fig:5ghz_Ha_pol}) is highly indicative that the Faraday rotation (and depolarization) is mainly occurring in the H\,$\alpha$ emitting shell around Coma~A, as opposed to the extended hot X-ray emitting gas that is also likely present \citep{Worrall2016}. Therefore, this makes a combination of an estimate of the electron density from the H\,$\alpha$ data, and the Faraday rotation information from the radio data a powerful tool in directly calculating the magnetic field strengths in this region. Here we follow the method outlined by \citet{McClure-Griffiths2010}. The electron density is related to the H\,$\alpha$ emission through the emission measure, $EM$:
    \begin{equation}
    \label{eq:EM_eqn}
    \mathrm{EM}= 2.75 \left(T / 10^4 \rm{K} \right)^{0.9} I_{\mathrm{H}\alpha} = n_e^2fL
    \end{equation}
    where $T$ is the temperature of the gaseous region in Kelvin, $I_{\mathrm{H}\alpha}$ is the H\,$\alpha$ intensity in Rayleighs (1 Rayleigh = 5.7 $\times 10^{-18}\,\rm{erg\,cm}^{-2}\,\rm{s}^{-1}\,\rm{arcsec}^{-2}$ at H\,$\alpha$), $n_e$ is the electron number density in cm$^{-3}$, $f$ is a filling factor between 0 and 1, and $L$ is the line of sight depth of the gaseous region in parsecs. Now using the Faraday rotation measure (Equation \ref{eq:Burn_RM_eqn}) and
     solving for the average values:
    \begin{equation}
    \label{eq:avg_B_eqn_temp}
    \left\langle B_\parallel \right\rangle = 1.235 \frac{\left\langle \mathrm{RM} \right\rangle }{\int n_e\left(l\right)\mathrm{d}l}
    \end{equation}
    
    It is reasonable to assume the main source of free electrons in the region between the source and observer is the H\,$\alpha$ region of thickness $L$. Expressing the column density of particles in this region along the line of sight as $n_efL=\sqrt{\mathrm{EM}fL}=\int n_e\left(l\right)\mathrm{d}l$, now we can rewrite Equation \ref{eq:avg_B_eqn_temp}:
    \begin{equation}
    \label{eq:avg_B_eqn}
    \left\langle B_\parallel \right\rangle = 1.235 \frac{\left\langle \mathrm{RM} \right\rangle }{n_efL}
    \end{equation}
    This gives an estimate of the average line of sight magnetic field in the ionised shell surrounding the radio lobes. 
    
     \par The $\sigma_{\mathrm{RM}}$ parameter from Section~\ref{subsec:RM_QU_theory} can be used to estimate the strength of the turbulent magnetic field component in the depolarizing shell of ionized gas. This can be calculated by considering the ionized gas as a collection of cells, each with their own Faraday depth and coherent magnetic field, which when combined along the line of sight gives a turbulent depolarizing screen. Using the definition of $\sigma_\mathrm{RM}$ given in \cite{Burn1966}, the diameter of such a cell, $d$ can be estimated as has been done in \cite{VanBreugel1984}: 
    \begin{equation}
    \sigma_\mathrm{RM} = 0.81 n_e \langle B_\text{rand} \rangle \sqrt{d L}
    \label{eq:sigma_Brand}
    \end{equation}
   
In Equation \ref{eq:sigma_Brand} all values except $B_\text{rand}$ and $d$ can be estimated directly from the data. A range of values for the turbulent cellsize can be estimated for each region using reasonable upper limits for $B_\text{rand}$ and $d$. 
The lower limit for the cellsize occurs when the value for the magnetic field strength is at maximum within each cell; a reasonable estimate for the maximum magnetic field is the equipartition field strength of the radio lobe in that region. This assumes the H\,$\alpha$ gas is completely mixed with the outer region of the lobe and depolarizes the signal, as well as disordering the direction of any magnetic field in the lobe in that region, but maintaining the same field strength in each of the cells of diameter $d$. The upper limit for $d$ can be taken to be half of the observing beam for the highest resolution observation of Coma A that shows depolarization. \cite{VanBreugel1985} have estimated the upper limit for $d$ using this method to be 300\,pc for all regions in Coma A, which yields the smallest possible $B_\text{rand}$ for the given depolarization. 

    For all our magnetic field calculations in Coma A we assumed $f=1$, $T=1\times 10^{4}$\,K, $L=3.5$\,kpc. The value used for $f$ assumes full ionisation in the gas. The temperature used here has been estimated in \cite{VanBreugel1985}. The thickness of the shell, $L$, was calculated by measuring the thickness of the projected shell in the H\,$\alpha$ image as seen in the filament of H\,$\alpha$ gas in the north east of the northern lobe (Figure~\ref{fig:archival_fpolHa}(b)). It must also be noted that the small Galactic contribution to the RM of $2.3 \pm 1.3$\,rad\,m$^{-2}$ \citep{Taylor2009} can be safely ignored due to the location of Coma A being very close to the Galactic pole.
 \subsubsection{Magnetic field strength in Faraday screen of northern lobe}   
    The northern hotspot region shows bright H\,$\alpha$ emission and some of the 2--4\,GHz polarization data for that region, particularly in the north western region of the hotspot is reliable for fitting (see Figure~\ref{fig:Sband_fit_images}). Values of $\mathrm{RM} \simeq 30\rm{rad\,m}^{-2}$, $\sigma_\mathrm{RM} \simeq  30\,\rm{rad\,m}^{-2}$, and $I_{\mathrm{H}\alpha} = 1.4 \times 10^{-16}\, $erg\,cm$^{-2}\,$s$^{-1}$arcsec$^{-2} \equiv 24.6\, \rm{Rayleigh} $ were used to estimate the magnetic field in the gas. The calculation also gives an estimate of $n_e=0.13\,\rm{cm}^{-3}$ for the electron density, as complete ionisation is assumed. This led to a value of $\left\langle B_\parallel \right\rangle = 0.07\,\umu$G for the ordered magnetic field in the ionised shell. The random component can be estimated to be $B_\text{rand} \geq 0.27\,\umu$G using the lower limit to the depolarizing cellsize of $d\leq300\,$pc. The smallest possible value for the cellsize $d$ is estimated to be $d\simeq 0.008\,$pc when the maximum possible magnetic field is assumed to be the equipartition estimate of 54\,$\umu$G \citep[Table 4. in ][]{VanBreugel1985} for that region. 
    
     The archival data can also be useful for estimating $\mathrm{RM}$ and $\sigma_\mathrm{RM}$, as the higher polarization fraction and range in $\lambda^2$ can give better estimates for the magnetic field strengths. Using the high frequency end of the 2--4\,GHz data and the 4.9\,GHz and 15\,GHz archival datasets, a simple linear fit of the polarization angles was used to estimate the $\mathrm{RM}$. To estimate $\sigma_\mathrm{RM}$, the fractional polarization for the same hotspot region averaged over a 2\arcsec~ beamsize was fit with the exponential part of the depolarization model (Equation \ref{eq:model_eqn}), namely:
    \begin{equation}
    p(\lambda) = p_0 \mathrm{e}^{-2\sigma_\mathrm{RM}^2\lambda^4}
    \label{eq:depol_fit}
    \end{equation}
    This method gave  $\mathrm{RM} \simeq  180\,\rm{rad\,m}^{-2}$ and $\sigma_\mathrm{RM} \simeq  176\,\rm{rad\,m}^{-2}$, which are significantly higher but may better describe the hotspot environment due to the improved resolution.  
    
    Using the rotation measure values and the same value for $I_{\mathrm{H}\alpha}$, the magnetic field strengths were calculated to be: $\left\langle B_\parallel \right\rangle = 0.45\,\umu$G and $\left\langle B_{\text{rand}} \right\rangle \geq 1.63\,\umu$G. The smallest possible value for the cellsize $d$ is estimated to be $d\simeq 0.5\,$pc when, the maximum possible magnetic field is again assumed to be the equipartition estimate of 54\,$\umu$G for the hotspot.  These values are almost an order of magnitude different from the earlier attempt, however the higher frequency data shows more reliable polarization structure. This estimate is supported by other studies \citep[e.g.][]{Laing2008, Guidetti2010, Bonafede2010} that commonly infer field strengths of 1 to 10 $\umu$G in the hot X-ray emitting gas phase that pervades the ambient intracluster/intragroup medium surrounding radio galaxies. In the case of Coma A, \cite{Worrall2016} argue that faint X-ray emission from thermal gas is also present, indicating a weak group atmosphere.
    
    The random and uniform line-of-sight magnetic field strength estimates of O(1)~$\umu$G suggest little or no mixing is occurring between the lobe and the surrounding gas near the northern hotspot. However, as the exact size of the depolarizing cellsize is unknown, the random magnetic field strength cannot be precisely estimated. In \cite{Mao2008} and \cite{Gaensler2005} a cell size of $\sim 90$\,pc is used for the Small and Large Magellanic Clouds. This value falls within the range of values estimated for $d$ and if the conditions in the Coma A H\,$\alpha$ region are similar to the  interstellar medium in the SMC and LMC the resulting  turbulent magnetic field can be estimated as $0.5\,\umu$G for the 2--4\,GHz results and $3\,\umu$G  for the results including the archival data at higher frequencies and resolution.
\subsubsection{Magnetic field strength in Faraday screen of southern lobe}
    Another interesting region to probe is the depolarizing `trough' feature transverse to the southern lobe discussed in Section \ref{subsec:structofComaA}. It is difficult to estimate RM and $\sigma_\mathrm{RM}$ in this region as even at 15\,GHz this region has no significant polarization. As the RM changes significantly in the jet downstream from this `trough', this H\,$\alpha$ filament may be causing this. Using a typical RM value for the downstream region, $\mathrm{RM} \simeq {-80}\,\rm{rad\,m}^{-2}$, the H\,$\alpha$ electron number density, $n_e \simeq 0.09\,\rm{cm}^{-3}$, and a shell depth of 3.5\,kpc, an estimate for the line-of-sight, coherent, magnetic field was calculated to be: $\left\langle B_\parallel \right\rangle = 0.31\,\umu$G. 
    
    Using the 15\,GHz archival data, an estimate of $\sigma_\mathrm{RM}$ can be made from the turbulent depolarization model. Assuming that the fractional polarization in regions upstream and downstream from the `trough' are very close to the intrinsic value at 15\,GHz, this implies that $p_{15\,\rm{GHz}} \simeq p_0 \simeq 0.3$ for these regions. This is a safe assumption as the model fitting has shown that $\sigma_\mathrm{RM}$ is very low in the inner lobe compared to the `trough' region. We also suppose that the `trough' region should have the same intrinsic polarization at zero wavelength as the regions on either side if not for strong depolarization by the H\,$\alpha$ gas. The fractional polarization in the `trough' region at 15\,GHz can be estimated to be less than  3\% polarized ($p(\lambda) \lesssim 0.03 $ ). 

Using Equation \ref{eq:depol_fit} with these assumptions, the resulting $\sigma_\mathrm{RM}$ value is $\sigma_\mathrm{RM} \geq  2700\,\rm{rad\,m}^{-2}$. This results in a random magnetic field strength of  $B_\text{rand} \geq 36\,\umu$G when using the upper limit for the depolarizing  cellsize of 300\,pc to give the lowest limit for this random magnetic field. This is comparable to the equipartition magnetic field strength across the southern lobe of 30 to 50$\,\umu$G \citep{VanBreugel1985}, and indicates that the H\,$\alpha$ gas has mixed with the synchrotron plasma of the lobes. Such strong evidence for internal Faraday rotation depolarization is rare but has been claimed in other recent studies \citep{OSullivan2013,Stawarz2013,Anderson2018}. There remains a possibility that the pre-existing magnetic field of O(1) $\mu$G (as estimated for the H\,$\alpha$ gas covering the northern lobe) is amplified significantly by a strong MHD shock, somewhat similar to that expected in supernova remnants \citep[e.g.][]{Schure2012}. However, it seems somewhat inconsistent that the necessary field amplification, by a factor of 30 or more, would occur only near the trough region, and not from the bow shock of the northern lobe also. 

In general, there are several potential reasons for the high depolarization. It may be due to a complex interaction between the lobe and the H\,$\alpha$ filament, where mixing between the filament and the lobe has occurred or the shock-ionisation of the filament mixed with the lobe has compressed it to result in such a high magnetic field strength. Furthermore this filament is spatially correlated with a decrease in the total intensity which can be seen in the contours in Figures~\ref{fig:Sband_Icont_spix} and \ref{fig:archival_fpolHa}(a), and in \cite{VanBreugel1985}; figure 5. is difficult to explain what is causing this decrease as a dense obstacle should ordinarily compress and enhance the synchrotron emission. However, one possibility given by \cite{VanBreugel1985} is that the interaction significantly increases the relative amount of the line-of-sight magnetic field component with respect to the perpendicular component at this location in the lobe. This would have the dual effect of decreasing the observed synchrotron emission and increasing internal Faraday rotation (and depolarization) at this location. 
This appears plausible as the magnitude of the RM is at its highest downstream of the filament, indicating a stronger line of sight magnetic field component and the depolarization is very high in the filament region with $\sigma_{\mathrm{RM}} \geq 2700\,\rm{rad\,m}^{-2} $.

\par Another possibility is that there is mass entrainment in the jet, shown by the region of low polarization along the jet centre in the southern lobe, and material has been `dumped' in the transverse `trough' region, which was proposed by \cite{VanBreugel1985}. This however is not an explanation for the high levels of H\,$\alpha$ emission in the `trough'/filament as no such emission is seen along the centre of the jet. The H\,$\alpha$ filaments also extend along the entire eastern side of Coma A including the northern lobe, where there is no evidence of mass entrainment. 
Therefore, we find that the most plausible scenario is that the H\,$\alpha$ gas shell is close to but external to the radio lobes, except in the `trough' region of the southern lobe, where the H\,$\alpha$ gas is mixing with the lobe plasma. 

\section{Conclusions}
We have presented a comprehensive broadband radio polarization and Faraday rotation study of the nearby radio galaxy Coma A. Broadband VLA observations in the 1--2\,GHz and 2--4\,GHz ranges were analysed using RM synthesis and qu-model fitting. Large variations in the observed degree of polarization were observed over the radio galaxy which is indicative of Faraday depolarization. Large variations in the RM were also observed across the southern lobe, indicating a substantial change in the morphology of the jet and lobe.  In conjunction with this, higher resolution archival observations at 4.9 and 15\,GHz were also analysed along with deep H\,$\alpha$ observations showing strong spatial correlation with the depolarized regions and the H\,$\alpha$ emission.

We find that an external Faraday depolarizing screen is consistent with the broadband polarization data across the majority of the radio lobes. Due to the clear anti-correlation between the polarized emission and H\,$\alpha$, this external Faraday screen is identified as the ionised gas seen in H\,$\alpha$ arcs and filaments surrounding the radio lobes. The H\,$\alpha$ emission is most likely caused by shock-ionisation of a gas disk (seen in neutral hydrogen) by the expanding radio lobes. 

\par We used the H\,$\alpha$ image in combination with the Faraday rotation and depolarization results to estimate the magnetic field strength in this ionised gas. We derived magnetic field strengths in the H\,$\alpha$ gas of $\sim 1\,\umu$G across most of the radio lobes, consistent with expectations for the magnetisation of gas in the local intergalactic medium. 
However, one particular H\,$\alpha$ filament that cuts across the southern lobe is clearly associated with a strong depolarization `trough', and has a magnetic field strength similar to the equipartition field strength in the lobes (i.e.~$\gtrsim 36\,\umu$G). This implies that the H\,$\alpha$ gas in this filament has mixed with the radio-emitting plasma of the lobes. This is one of the clearest direct observational examples of internal Faraday depolarization in the lobe of a radio galaxy. This adds to the complex ways that radio jets and lobes can interact with their environment, providing another means of ``feedback'' on local gas, that can both inhibit further star formation activity and substantially modify the evolution of radio galaxies.

\section*{Acknowledgements}

This research has been funded by the Irish Research Council under the Government of Ireland Postgraduate Programme. The VLA is an instrument of the National Radio Astronomy Observatory, a facility of the National Science Foundation operated under cooperative agreement by Associated Universities, Inc. The authors would like to thank Clive Tadhunter for the H\,$\alpha$ image of Coma A and Denise Gabuzda for useful discussion and feedback during the writing of this manuscript. The authors also wish to thank the referee, Dr. J. P. Leahy, for his thoughtful feedback which improved this paper.  SPO acknowledges financial support from the Deutsche Forschungsgemeinschaft (DFG) under grant BR2026/23 and from UNAM through the PAPIIT project IA103416.




\bibliographystyle{mnras.bst}

\bibliography{ComaAwork} 

\begin{thebibliography}{}
\makeatletter
\relax
\def\mn@urlcharsother{\let\do\@makeother \do\$\do\&\do\#\do\^\do\_\do\%\do\~}
\def\mn@doi{\begingroup\mn@urlcharsother \@ifnextchar [ {\mn@doi@}
  {\mn@doi@[]}}
\def\mn@doi@[#1]#2{\def\@tempa{#1}\ifx\@tempa\@empty \href
  {http://dx.doi.org/#2} {doi:#2}\else \href {http://dx.doi.org/#2} {#1}\fi
  \endgroup}
\def\mn@eprint#1#2{\mn@eprint@#1:#2::\@nil}
\def\mn@eprint@arXiv#1{\href {http://arxiv.org/abs/#1} {{\tt arXiv:#1}}}
\def\mn@eprint@dblp#1{\href {http://dblp.uni-trier.de/rec/bibtex/#1.xml}
  {dblp:#1}}
\def\mn@eprint@#1:#2:#3:#4\@nil{\def\@tempa {#1}\def\@tempb {#2}\def\@tempc
  {#3}\ifx \@tempc \@empty \let \@tempc \@tempb \let \@tempb \@tempa \fi \ifx
  \@tempb \@empty \def\@tempb {arXiv}\fi \@ifundefined
  {mn@eprint@\@tempb}{\@tempb:\@tempc}{\expandafter \expandafter \csname
  mn@eprint@\@tempb\endcsname \expandafter{\@tempc}}}

\bibitem[\protect\citeauthoryear{Anderson, Gaensler, Heald, O'Sullivan,
  Kaczmarek  \& Feain}{Anderson et~al.}{2018}]{Anderson2018}
Anderson C.~S.,  Gaensler B.~M.,  Heald G.~H.,  O'Sullivan S.~P.,  Kaczmarek
  J.~F.,   Feain I.~J.,  2018, \mn@doi [\apj] {10.3847/1538-4357/aaaec0},
  855:41, 27

\bibitem[\protect\citeauthoryear{Bonafede, Feretti, Murgia, Govoni, Giovannini,
  Dallacasa, Dolag  \& Taylor}{Bonafede et~al.}{2010}]{Bonafede2010}
Bonafede A.,  Feretti L.,  Murgia M.,  Govoni F.,  Giovannini G.,  Dallacasa
  D.,  Dolag K.,   Taylor G.~B.,  2010, \mn@doi [A{\&}A]
  {10.1051/0004-6361/200913696}, 513, A30

\bibitem[\protect\citeauthoryear{Brentjens \& de Bruyn}{Brentjens \&
  de~Bruyn}{2005}]{Brentjens2005}
Brentjens M.~A.,  de Bruyn a.~G.,  2005, \mn@doi [A{\&}A]
  {10.1051/0004-6361:20052990}, 441, 1217

\bibitem[\protect\citeauthoryear{Bridle, Fomalont, Palimaka  \& Willis}{Bridle
  et~al.}{1981}]{Bridle1981}
Bridle A.~H.,  Fomalont E.~B.,  Palimaka J.~J.,   Willis A.~G.,  1981, \mn@doi
  [ApJ] {10.1086/159174}, 248, 499

\bibitem[\protect\citeauthoryear{Burn}{Burn}{1966}]{Burn1966}
Burn B.~J.,  1966, \mn@doi [MNRAS] {10.1093/mnras/133.1.67}, 133, 67

\bibitem[\protect\citeauthoryear{Cant{\'{o}}, Curiel  \&
  Mart{\'{i}}nez-G{\'{o}}mez}{Cant{\'{o}} et~al.}{2009}]{Canto2009}
Cant{\'{o}} J.,  Curiel S.,   Mart{\'{i}}nez-G{\'{o}}mez E.,  2009, \mn@doi
  [A{\&}A] {10.1051/0004-6361/200911740}, 501, 1259

\bibitem[\protect\citeauthoryear{Chiaberge, Capetti  \& Celotti}{Chiaberge
  et~al.}{1999}]{Chiaberge1999}
Chiaberge M.,  Capetti A.,   Celotti A.,  1999, A{\&}A, 349, 77

\bibitem[\protect\citeauthoryear{Fabian}{Fabian}{2012}]{Fabian2012}
Fabian A.~C.,  2012, \mn@doi [\araa] {10.1146/annurev-astro-081811-125521}, 50,
  455

\bibitem[\protect\citeauthoryear{Fanaroff \& Riley}{Fanaroff \&
  Riley}{1974}]{Fanaroff1974}
Fanaroff B.~L.,  Riley J.~M.,  1974, \mn@doi [MNRAS] {10.1093/mnras/167.1.31P},
  167, 31P

\bibitem[\protect\citeauthoryear{Gaensler}{Gaensler}{2005}]{Gaensler2005}
Gaensler B.~M.,  2005, \mn@doi [Science] {10.1126/science.1108832}, 307, 1610

\bibitem[\protect\citeauthoryear{Garrington, Leahy, Conway  \&
  Laingt}{Garrington et~al.}{1988}]{Garrington1988}
Garrington S.~T.,  Leahy J.~P.,  Conway R.~G.,   Laingt R.~A.,  1988, \mn@doi
  [Nature] {10.1038/331147a0}, 331, 147

\bibitem[\protect\citeauthoryear{Gawronski, Marecki, Kunert-Bajraszewska  \&
  Kus}{Gawronski et~al.}{2006}]{Gawronski2006}
Gawronski M.~P.,  Marecki A.,  Kunert-Bajraszewska M.,   Kus A.~J.,  2006,
  \mn@doi [A{\&}A] {10.1051/0004-6361/201220544}, 447, 63

\bibitem[\protect\citeauthoryear{Gopal-Krishna \& Wiita}{Gopal-Krishna \&
  Wiita}{2000a}]{Gopal-Krishna2000b}
Gopal-Krishna Wiita P.~J.,  2000a, A{\&}A, 363, 507

\bibitem[\protect\citeauthoryear{{Gopal-Krishna} \& {Wiita}}{{Gopal-Krishna} \&
  {Wiita}}{2000b}]{Gopal-Krishna2000a}
{Gopal-Krishna} {Wiita} P.~J.,  2000b, \mn@doi [\apj] {10.1086/308230}, \href
  {http://adsabs.harvard.edu/abs/2000ApJ...529..189G} {529, 189}

\bibitem[\protect\citeauthoryear{Guidetti, Laing, Murgia, Govoni, Gregorini  \&
  Parma}{Guidetti et~al.}{2010}]{Guidetti2010}
Guidetti D.,  Laing R.~A.,  Murgia M.,  Govoni F.,  Gregorini L.,   Parma P.,
  2010, \mn@doi [A{\&}A] {10.1051/0004-6361/200913872}, 514, 50

\bibitem[\protect\citeauthoryear{Heckman \& Best}{Heckman \&
  Best}{2014}]{Heckman2014}
Heckman T.,  Best P.,  2014, \mn@doi [\araa]
  {10.1146/annurev-astro-081913-035722}, 52, 589

\bibitem[\protect\citeauthoryear{{Knuettel}, {Gabuzda}  \&
  {O'Sullivan}}{{Knuettel} et~al.}{2017}]{knuettel2017}
{Knuettel} S.,  {Gabuzda} D.,   {O'Sullivan} S.,  2017, \mn@doi [Galaxies]
  {10.3390/galaxies5040061}, \href
  {http://adsabs.harvard.edu/abs/2017Galax...5...61K} {5, 61}

\bibitem[\protect\citeauthoryear{{Laing}}{{Laing}}{1988}]{Laing1988}
{Laing} R.~A.,  1988, \mn@doi [\nat] {10.1038/331149a0}, \href
  {http://adsabs.harvard.edu/abs/1988Natur.331..149L} {331, 149}

\bibitem[\protect\citeauthoryear{Laing, Bridle, Parma  \& Murgia}{Laing
  et~al.}{2008}]{Laing2008}
Laing R.~A.,  Bridle A.~H.,  Parma P.,   Murgia M.,  2008, \mn@doi [MNRAS]
  {10.1111/j.1365-2966.2008.13895.x}, 391, 521

\bibitem[\protect\citeauthoryear{Leahy \& Fernini}{Leahy \&
  Fernini}{1989}]{Leahy1989}
Leahy P.,  Fernini I.,  1989, Technical report, {VLA Scientific Memorandum No.
  161 Correction Schemes for Polarized Intensity}.
National Radio Astronomy Observatory

\bibitem[\protect\citeauthoryear{Mao, Gaensler, Stanimirovi{\'{c}}, Haverkorn,
  McClure-Griffiths, Staveley-Smith  \& Dickey}{Mao et~al.}{2008}]{Mao2008}
Mao S.~A.,  Gaensler B.~M.,  Stanimirovi{\'{c}} S.,  Haverkorn M.,
  McClure-Griffiths N.~M.,  Staveley-Smith L.,   Dickey J.~M.,  2008, \mn@doi
  [\apj] {10.1086/590546}, 688, 1029

\bibitem[\protect\citeauthoryear{McClure-Griffiths, Madsen, Gaensler, McConnell
   \& Schnitzeler}{McClure-Griffiths et~al.}{2010}]{McClure-Griffiths2010}
McClure-Griffiths N.~M.,  Madsen G.~J.,  Gaensler B.~M.,  McConnell D.,
  Schnitzeler D. H. F.~M.,  2010, \mn@doi [ApJ] {10.1088/0004-637X/725/1/275},
  725, 275

\bibitem[\protect\citeauthoryear{Morganti, Oosterloo, Tinti, Tadhunter, Wills
  \& van Moorsel}{Morganti et~al.}{2002}]{Morganti2002}
Morganti R.,  Oosterloo T.~A.,  Tinti S.,  Tadhunter C.~N.,  Wills K.~A.,   van
  Moorsel G.,  2002, \mn@doi [A{\&}A] {10.1051/0004-6361:20020350}, 387, 830

\bibitem[\protect\citeauthoryear{O'Sullivan et~al.,}{O'Sullivan
  et~al.}{2012}]{OSullivan2012}
O'Sullivan S.~P.,  et~al., 2012, \mn@doi [MNRAS]
  {10.1111/j.1365-2966.2012.20554.x}, 421, 3300

\bibitem[\protect\citeauthoryear{O'Sullivan et~al.,}{O'Sullivan
  et~al.}{2013}]{OSullivan2013}
O'Sullivan S.~P.,  et~al., 2013, \mn@doi [\apj] {10.1088/0004-637X/764/2/162},
  764, 162

\bibitem[\protect\citeauthoryear{{Planck Collaboration} et~al.,}{{Planck
  Collaboration} et~al.}{2016}]{Planckcoll2016}
{Planck Collaboration} et~al., 2016, \mn@doi [A{\&}A]
  {10.1051/0004-6361/201526926}, 594, A11

\bibitem[\protect\citeauthoryear{{Schure}, {Bell}, {O'C Drury}  \&
  {Bykov}}{{Schure} et~al.}{2012}]{Schure2012}
{Schure} K.~M.,  {Bell} A.~R.,  {O'C Drury} L.,   {Bykov} A.~M.,  2012, \mn@doi
  [\ssr] {10.1007/s11214-012-9871-7}, \href
  {http://adsabs.harvard.edu/abs/2012SSRv..173..491S} {173, 491}

\bibitem[\protect\citeauthoryear{Sokoloff, Bykov, Shukurov, Berkhuijsen, Beck
  \& Poezd}{Sokoloff et~al.}{1998}]{sokoloff1998}
Sokoloff D.~D.,  Bykov A.~A.,  Shukurov A.,  Berkhuijsen E.~M.,  Beck R.,
  Poezd A.~D.,  1998, \mn@doi [MNRAS] {10.1046/j.1365-8711.1998.01782.x}, 299,
  189

\bibitem[\protect\citeauthoryear{Stawarz et~al.,}{Stawarz
  et~al.}{2013}]{Stawarz2013}
Stawarz et~al., 2013, \mn@doi [\apj] {10.1088/0004-637X/766/1/48}, 766, 48

\bibitem[\protect\citeauthoryear{Tadhunter, Villar-Martin, Morganti,
  Bland-Hawthorn  \& Axon}{Tadhunter et~al.}{2000}]{Tadhunter2000}
Tadhunter C.~N.,  Villar-Martin M.,  Morganti R.,  Bland-Hawthorn J.,   Axon
  D.,  2000, \mn@doi [MNRAS] {10.1046/j.1365-8711.2000.03416.x}, 314, 849

\bibitem[\protect\citeauthoryear{Taylor, Stil  \& Sunstrum}{Taylor
  et~al.}{2009}]{Taylor2009}
Taylor A.~R.,  Stil J.~M.,   Sunstrum C.,  2009, \mn@doi [ApJ]
  {10.1088/0004-637X/702/2/1230}, 702, 1230

\bibitem[\protect\citeauthoryear{Worrall, Birkinshaw  \& Young}{Worrall
  et~al.}{2016}]{Worrall2016}
Worrall D.~M.,  Birkinshaw M.,   Young A.~J.,  2016, \mn@doi [MNRAS]
  {10.1093/mnras/stw277}, 458, 174

\bibitem[\protect\citeauthoryear{Wright}{Wright}{2006}]{Wright2006}
Wright E.~L.,  2006, \mn@doi [\pasp] {10.1086/510102}, 118, 1711

\bibitem[\protect\citeauthoryear{de
  Gasperin}{de~Gasperin}{2017}]{DeGasperin2017}
de Gasperin F.,  2017, \mn@doi [MNRAS] {10.1093/mnras/stx210}, 467, 2234

\bibitem[\protect\citeauthoryear{van Breugel, Heckman  \& Miley}{van Breugel
  et~al.}{1984}]{VanBreugel1984}
van Breugel W.,  Heckman T.,   Miley G.,  1984, \mn@doi [\apj]
  {10.1086/161594}, 276, 79

\bibitem[\protect\citeauthoryear{van Breugel, Miley, Heckman, Butcher  \&
  Bridle}{van Breugel et~al.}{1985}]{VanBreugel1985}
van Breugel W.,  Miley G.,  Heckman T.,  Butcher H.,   Bridle A.,  1985,
  \mn@doi [ApJ] {10.1086/163007}, 290, 496

\makeatother
\end{thebibliography}





\bsp	
\label{lastpage}
\end{document}